\documentclass[twocolumn,showpacs,amsmath,amssymb,aps]{revtex4}

\usepackage{graphicx}
\usepackage{psfrag}
\usepackage{dcolumn}
\usepackage{bm}

\begin{document}


\title{Non-Newtonian thin films with normal stresses: dynamics and spreading.}

\author{Arezki Boudaoud}
\email{arezki.boudaoud@lps.ens.fr}
\affiliation{%
Laboratoire de Physique Statistique, 
\'Ecole normale sup\'erieure, 24 rue Lhomond, F-75231 PARIS Cedex 05, France}%

\date{\today}

\begin{abstract}
The dynamics of thin films on a horizontal solid substrate is investigated in the case of non-Newtonian fluids exhibiting normal stress differences, the rheology of which is strongly non-linear. Two coupled equations of evolution for the thickness of the film and the shear rate are proposed within the lubrication approximation. This framework is applied to the motion of an advancing contact line.  The apparent dynamic contact angle is found to depend logarithmically on a lengthscale determined solely by the rheological properties of the fluid and the velocity of the contact line.
\end{abstract}


\maketitle

The spreading of a thin fluid layer on a substrate has received much attention due to its practical importance. However, the motion of a contact line is still a matter of debate  (see Refs.~\cite{dussan,degennes,pomeau,eggers} for a review). Macroscopically, there is a balance between viscous forces (shear viscosity $\mu$) and capillary forces (surface tension $\sigma$). This results in the Cox-Voinov law~\cite{cox}, which relates the apparent (or dynamic) contact angle $\theta_\mathrm{d}$ to the velocity $U$ of the contact line 
\begin{equation}
\theta_\mathrm{d}^3=9\frac{\mu U}{\gamma} \ln (x/\ell_\mathrm{m}),\label{intr1}
\end{equation}
$x$ being the distance to the contact line. This equation is ill-defined for small  $x$ which reflects the divergence of the viscous stresses at the contac-line~\cite{huh}. The value of the  length $\ell_\mathrm{m}$ depends on the  regularising microscopic physics accounted for in the model --- e.g. Van der Waals forces~\cite{degennes}, slip~\cite{huh} or diffuse interface~\cite{pomeau} --- so that macroscopic measurements can be used to probe microscopic properties. Experiments on the spreading of silicon oils~\cite{kavehpour} confirm the model based on Van der Waals forces~\cite{degennes}. However it is plausible that the relevant model depends on the nature of both the fluid and the substrate.

In applications, most fluids are complex and exhibit non-Newtonian properties. Except for some viscoelastic fluids~\cite{rauscher},  they have a nonlinear constitutive equation, which raises a theoretical challenge~\cite{rosenblat}. Until now, lubrication theories were restricted to fluids with no normal stresses, such as yield-stress fluids~\cite{balmforth} or shear-thinning fluids~\cite{gorodtsov,king,king2,starov,neogi,betelu03,rafai,betelu04}. Shear-thinning was even proposed as the regularising microscopical mechanism~\cite{weidner,ansini,carre02}. Experimental studies are fewer~\cite{carre97,carre00,rafai}; the more recent one~\cite{rafai} also considered fluids for which the only non-Newtonian property  is the existence of normal stresses, for which no theoretical framework was available.

In this Letter we consider the spreading of a thin layer of fluid having a constant shear viscosity $\mu$  and exhibiting first normal stresses difference~\cite{bird} $\sigma_{xx}-\sigma_{zz}=\psi (\partial_z v)^2$, $\sigma$ being the stress tensor (see Fig.~1 for the geometry and other notations). In dilute polymeric suspensions, the second normal stress difference $\sigma_{yy}-\sigma_{zz}$ is negligible and the normal stress coefficient $\psi$ can be considered as constant~\cite{bird}.
Within the lubrication approximation, we propose a set of two coupled equations of evolution (Eqs.~\ref{pde1D}-\ref{pde2D}) for the film height and the shear rate (averaged over the thickness). Then we investigate an advancing contact line (at velocity $U$). In particular we determine the lengthscale which replaces the microscopic length in~(\ref{intr1}), 
\begin{equation}
\ell_\mathrm{m}=\frac{\psi U}{b \mu},\label{intr2}
\end{equation}
as a result from the existence of normal stresses, $b$ being a numerical constant. Thus we give more grounds to the scaling analysis of Ref.~\cite{rafai}.

\begin{figure}
    \centering
    \includegraphics[width=.9\columnwidth]{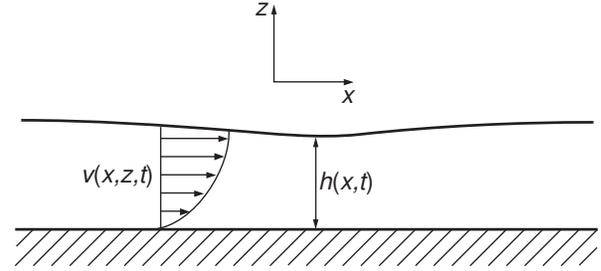}
    \caption{Schematic of the geometry, defining the directions $x,z$, the film thickness $h$ and horizontal velocity $v$.}
    \label{schem}
\end{figure}

In view of the lubrication approximation, we introduce the aspect ratio $\epsilon=Z/X$ of the film, $Z$ being a typical thickness and $X$ the horizontal lengthscale. Let $p$ be the pressure field and $\phi$ the potential of an applied body force such as a microscopic force or gravity (then $\phi=\rho g z$, $\rho$ being the volumic mass). The stress balance reads
\begin{eqnarray}
\partial_x\sigma_{xx}+\partial_z\sigma_{zx}=\partial_x \phi+\partial_x p\label{eq1}\\
\partial_x\sigma_{xz}+\partial_z\sigma_{zz}=\partial_z \phi+\partial_z p\label{eq2}
\end{eqnarray}
As $\partial_x/\partial_z\sim\epsilon\ll 1$,  Eq.~(\ref{eq2}) yields at the lower order in $\epsilon$ that 
$\pi=p+\phi-\sigma_{zz}$ is a function of $x$ only. Then Eq. (\ref{eq1}) becomes $\partial_x(\sigma_{xx}-\sigma_{zz})+\partial_z\sigma_{zx}=\partial_x \pi$, i.e. , using the rheology,
\begin{equation}
\mu\partial_{zz}v+\psi \ \partial_x\left[(\partial_z v)^2\right]=\partial_x \pi.\label{eqq}
\end{equation}
and $\pi$ is determined using the normal stress balance $-p+\sigma_{zz}=\gamma \kappa$ at the free surface, accounting for its surface tension $\gamma$ and curvature $\kappa$, so that
\begin{equation}
\pi=-\gamma \partial_{zz} h +\phi(z=h).\label{press}
\end{equation}
The set (\ref{eqq}-\ref{press}) is closed with mass conservation
\begin{equation}
\partial_t h+\partial_x \int_0^h v(z,t)\mathrm{d}z=0.\label{cons}
\end{equation}
Thus we obtain a system of PDEs for $h$ and $v$.
In fact, Eq.~(\ref{eqq}) can be solved for $v$ by a series of the form
\begin{equation}
v(x,z,t)=\Sigma_{n=0}^{\infty}\ a_n(x)z^n, \label{series}
\end{equation}
where $a_0=0$ to ensure no slip at the substrate, $a_1(x,t)=2s(x,t)$ is proportional to the mean shear rate across the thickness, $a_2=\partial_x(- \pi +4 \psi s^2)/\mu$, and each following $a_n$ can be computed recursively with the $x$-derivatives of the previous coefficients.  Here we propose to truncate the expansion at order 2: $v=2 s z + a_2 z^2$. This truncation is further discussed in the conclusion, but it can be noted beforehand that it is obviously exact in the standard case of no normal stress as well as in the case of strong normal stress where Eq.~(\ref{eqq}) shows that the velocity profile is linear in $z$. Then the condition of no shear stress at the free surface $\partial_z v=0$ yields a second equation relating $a_2$ to $s$.  As a consequence Eqs.~(\ref{eqq}--\ref{cons}) reduce to two coupled PDEs for the thickness $h$ and the mean shear rate $s$,
\begin{eqnarray}
2 \mu s+h\, \partial_x\left(\pi -4\psi s^2\right)=0,\label{pde1}\\
\partial_t h+\frac{2}{3}\partial_x \left(h^2 s\right)=0.\label{pde2}
\end{eqnarray}
This set is readily generalised to account for a third direction $y$; $\mathbf{s}=(s_x,s_y)$ is then the vectorial mean shear:
\begin{eqnarray}
2 \mu \mathbf{s}+h\, \mathbf{\nabla}\left(\pi -4 \psi \mathbf{s}^2\right)=0,\label{pde1D}\\
\partial_t h+\frac{2}{3}\mathbf{\nabla\cdot} \left(h^2 \mathbf{s}\right)=0,\label{pde2D}
\end{eqnarray}
where the dynamic pressure $\pi$ is defined by Eq.~\ref{press} and $\mathbf{\nabla}=(\partial_x,\partial_y)$.

Now we proceed to the study of a moving contact line, the only driving forces being the capillary forces:
$\pi=-\gamma \partial_{xx}h$. We consider a contact line advancing at constant velocity $U$ towards $x=-\infty$, i.e. $h(x,t)=h(x+Ut)$ and $s(x,t)=s(x+Ut)$ ,
so that Eq.~(\ref{pde2}) reads
$
U+2/3\ h s=0.
$
Replacing $s$ in \ref{pde2} yields
\begin{equation}
1 +\frac{1}{3\mathcal{C}} h^2 h'''-6\ell \frac{h'}{h}=0.\label{cline}
\end{equation}
where $\mathcal{C}=\mu U/\gamma$ is the capillary number and $\ell=\psi U/\mu$ is the normal stress characteristic length. The scaling form of the solutions to (\ref{cline}) is
\begin{equation}
x=X\ell, \quad  h(x)=\mathcal{C}^{1/3} \ell\, H(X),
\end{equation}
which yields 
\begin{equation}
1 +\frac{1}{3} H^2 H'''-6 \frac{H'}{H}=0.\label{clinad}
\end{equation}
We look for the solutions of Eq.~(\ref{clinad}) which vanish at $X=0$ and have no macroscopic curvature, i.e. $H''\to 0$ when $X\to+\infty$. Such a solution has the expansion $H(X)=(3/2)^{1/3}X^{2/3}(3+X+a X^2 + O(X^3))$ near $X=0$. Shooting on the value of $a$  yields the only solution with no macroscopic curvature. It has the classical asymptotic form $H=3^{2/3}X[\ln(bX)]^{1/3}$, with $b=1.69$. This solution is depicted in Fig.~2 and allows a matching between Cox-Voinov's law (\ref{intr1}) and the $H\sim X^{2/3}$ scaling resulting from the balance between capillarity and normal stresses near the contact line.

\begin{figure}
    \centering
    \includegraphics[width=.8\columnwidth]{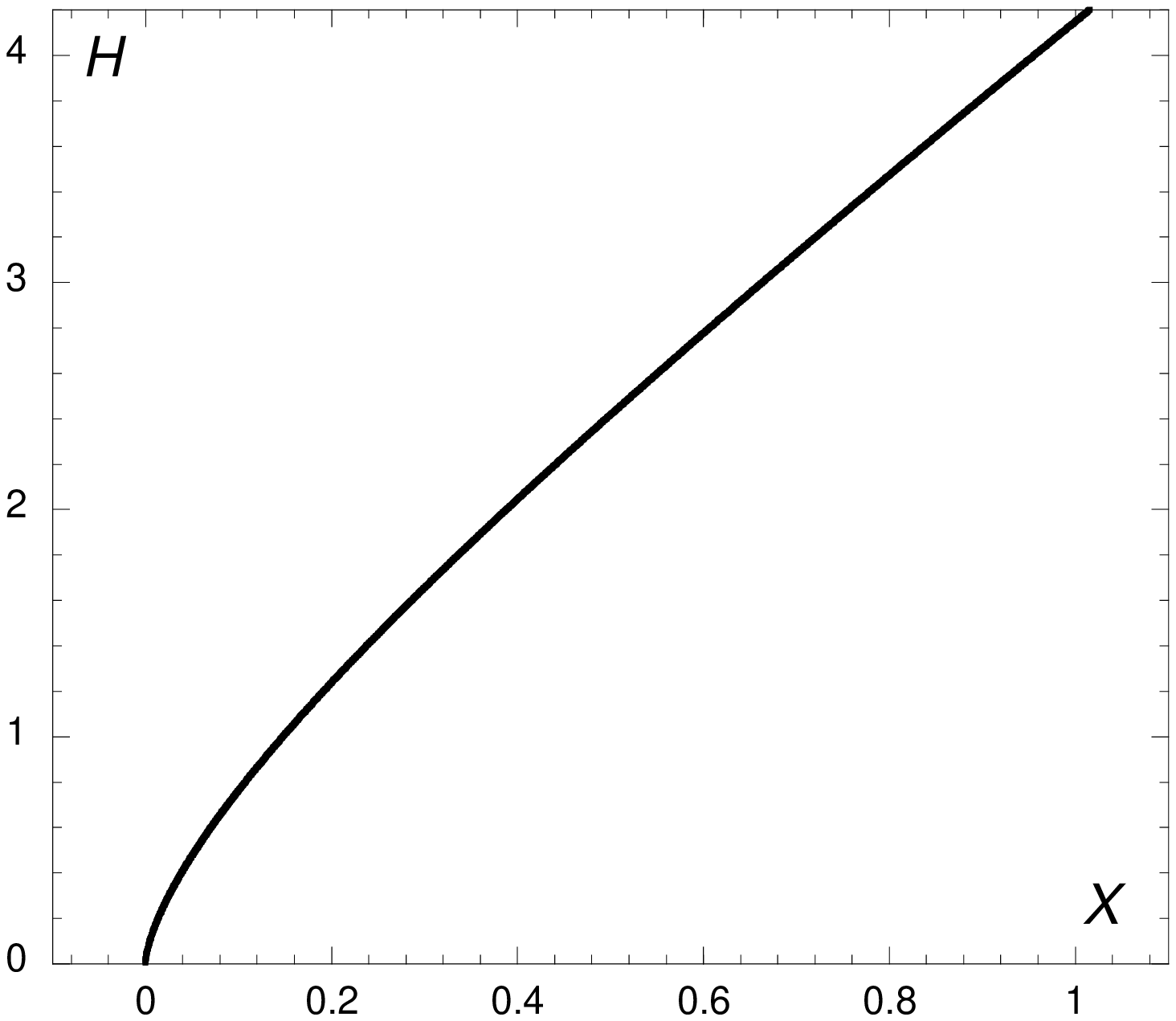}
    \includegraphics[width=.8\columnwidth]{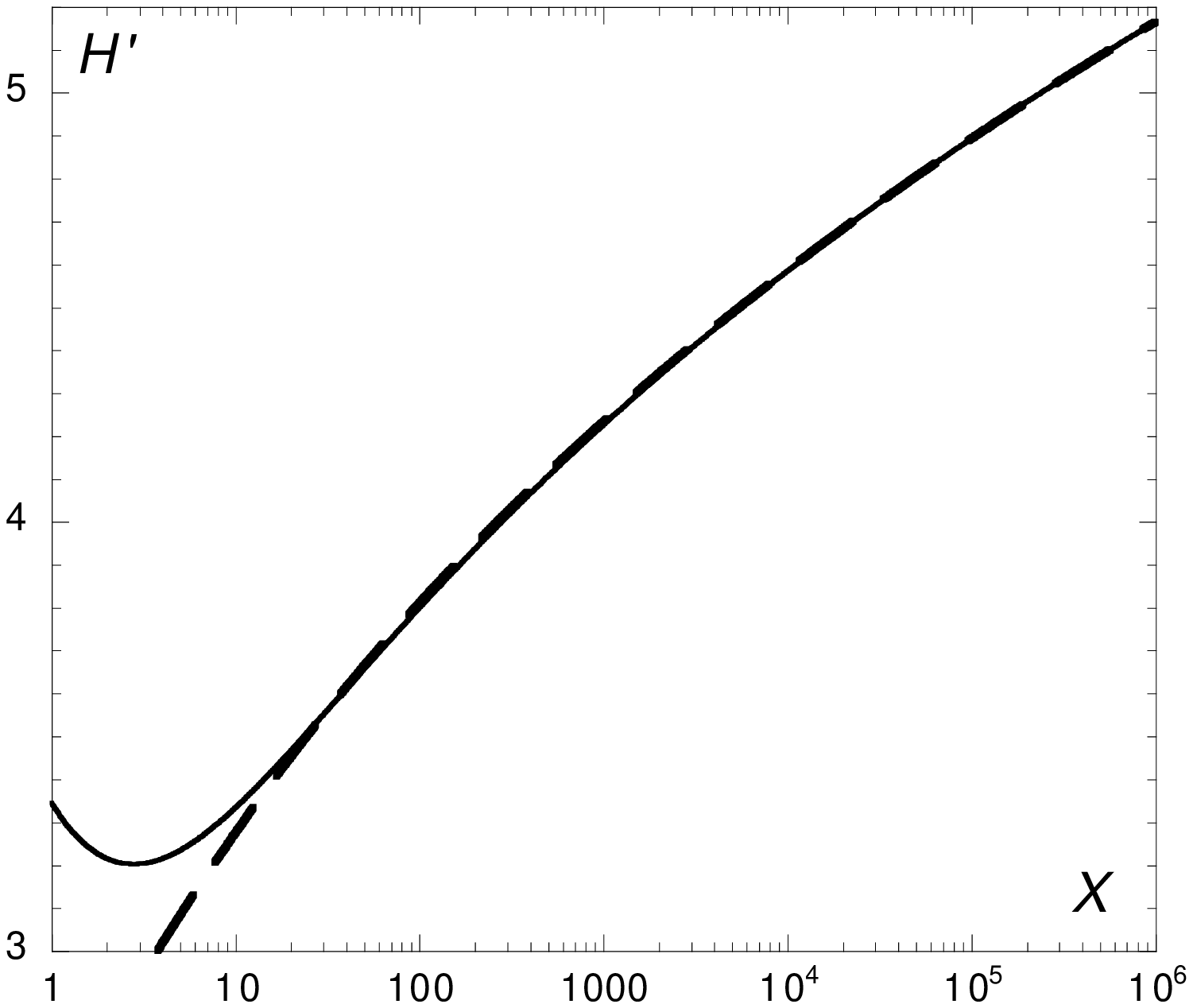}
    \caption{Film thickness for an advancing contact line as given by the solution to (\ref{clinad}). \textbf{a} Thickness $H(X)$. \textbf{b} Slope $H'(X)$ (continuous line) and comparison with the asymptotic form $\left(9 \ln(bX)\right)^{1/3}$  with $b=1.69$ (dashed line).}
    \label{schem}
\end{figure}

To summarise, we proposed the set (\ref{pde1D}-\ref{pde2D}) of coupled PDEs for the film thickness and mean shear. It was obtained using a truncation which is exact in both the limits of no and strong normal stresses. Within this framework, we showed that the rheology provides a regularising lengthscale $\ell_\mathrm{m}$ (Eq.~\ref{intr2}) which is of the order of 1$\mu$m in experiments~\cite{rafai}. Obviously the results are valid as long as  $\ell_\mathrm{m}$ is much larger than any microscopic length such as a slip length or the size of a precursor film. Here, the divergence of the viscous dissipation $\mathcal{D}\sim\int\mathrm{d}X/H$ is removed by the $H\sim X^{2/3}$ scaling of the film thickness near the contact line.
The present study could be improved by a truncation at higher order, although the robust asymptotics very near to and far from the contact line would not be altered; however this would yield a formidable numerical task as the order of the PDEs would increase with the order of the truncation. Another extension is to match the region where normal stresses balance capillarity to the smaller region where microscopic physics become important.

I am grateful to M. Ben Amar, D. Bonn and S. Rafai for introducing me to the spreading of non-Newtonian fluids and for fruitful discussions. Laboratoire de Physique Statistique is UMR 8550 of  CNRS and is associated with ENS and the universities of Paris VI and VII.

\bibliography{nonnewt}

\end{document}